\documentclass{article}

\usepackage[english]{babel}

\usepackage[letterpaper,top=2cm,bottom=2cm,left=3cm,right=3cm,marginparwidth=1.75cm]{geometry}

\usepackage{amsmath}
\usepackage{graphicx}
\usepackage[colorlinks=true, allcolors=blue]{hyperref}

\title{Scaling the human niche\footnote{
Peer-reviewed chapter in press for an edited volume \textit{Scaling in Biology: A New Synthesis}, Editors: B.J. Enquist, C. Kempes and M. O’Connor. Santa Fe Institute Press.}}

\author{Marcus J. Hamilton
\\
\\
\normalsize{Department of Anthropology,} \\
\normalsize{University of Texas at San Antonio,}\\ 
\normalsize{San Antonio, TX 78249, USA.}\\
\\
\normalsize{School of Data Science,} \\
\normalsize{University of Texas at San Antonio,}\\ 
\normalsize{San Antonio, TX 78207, USA.}\\
\\
\normalsize{Santa Fe Institute,} \\
\normalsize{1399 Hyde Park Rd,}\\ 
\normalsize{Santa Fe, NM 87501, USA.}\\
\\
\normalsize{E-mail: marcus.hamilton@utsa.edu}
}

\begin{document}
\maketitle

\begin{quote}
    \textit{``The concept of the niche provides a way of characterizing important ecological attributes of species while recognizing their uniqueness.''}
\hspace*{\fill}James H. Brown \cite{brown1995macroecology}\quad
\end{quote}

\begin{quote}
    \textit{``Culture is viewed as the extrasomatic means of adaptations for the human organism (White 1959:8).''}
\hspace*{\fill}Lewis R. Binford \cite{binford1962archaeology} citing Leslie A. White \cite{white2016evolution}\quad
\end{quote}

\begin{abstract}
    The human niche represents the intersection of biological, ecological, cultural, and technological processes that have co-evolved to shape human adaptation and societal complexity. This paper explores the human niche through the lens of macroecological scaling theory, seeking to define and quantify the dimensions along which human ecological strategies have diversified. By leveraging concepts from classic niche theory, niche construction, and complex adaptive systems, I develop a framework for understanding human ecology as both predictable within mammalian scaling relationships and uniquely divergent due to social, cognitive, and technological factors. Key dimensions of the human niche—metabolism, cognition, sociality, and computation—are examined through scaling laws that structure human interactions with the environment and each other. The paper demonstrates how human niche expansion over evolutionary time has been characterized by increasing metabolic consumption, information processing capacity, and the formation of larger, more interconnected social networks. This cumulative trajectory has led to the unprecedented scale of contemporary human societies, with implications for sustainability, economic development, and future niche expansion, including into space. The study underscores the need for an integrative, quantitative approach to human ecology that situates human adaptability within broader ecological and evolutionary constraints.
\end{abstract}

\section{Introduction}

The concept of the human niche is used in anthropology to refer to a coevolved suite of genetic, cultural, and ecological adaptations that differentiate human lifestyles (and those of our hominin ancestors) from other species. James Brown's quote above from his monograph \textit{Macroecology} \cite{brown1995macroecology} captures the ability of the niche concept to simultaneously juxtapose the idiosyncrasies of a focal species while referring to the general set of ecological characteristics shared by all species. Lewis Binford's quote from his influential paper \textit{Archaeology as Anthropology} \cite{binford1962archaeology} cites a well-known dictum from Leslie White who had defined culture a few years before as the set of behaviors, technologies, and norms that humans use to ``sustain and perpetuate...existence'' \cite{white2016evolution}. Although neither White nor Binford used the term niche explicitly at this time (Binford later would), they capture both the evolutionary and ecological characteristics of the concept and it's logical extension into the human species, and it is this they define as ``culture''. 

White, Binford and others in the 1960s laid the foundations of modern archaeological science while parallel developments in cultural anthropology, influenced by models from ecology and economics, would lead to the field of human behavioral ecology. Similar conceptual developments in ecology and complex systems science would lead to macroecology, biological scaling, and complex adaptive systems theory. However, despite convergent interests, for historical reasons  ecology, anthropology, archaeology, and complexity science never fully integrated theoretically and so while concepts may be shared across closely related domains, their use can be poorly defined. The concept of the niche is a case in point.

In this chapter, I show how the concept of human niche can be defined using macroecological scaling theory. Effectively, here I explore the ways in which the ecology of the human species is predictable for a mammal, and the ways in which it is unique. Understanding how human ecology deviates from expectations is ultimately a statistical question that requires empirical data, statistical models, and mathematical theory. Specifically, I develop a macroecological approach that uses models derived from ecological theory to set a baseline--or a frame of reference \cite{binford2019constructing}--from which we can measure deviations from expectations. Identifying these deviations then highlights the specific ecological and evolutionary trajectories along which human ecology has diversified.

\section{What is the human niche?}

\begin{figure}[ht!]
\centering
\includegraphics[width=1\textwidth]{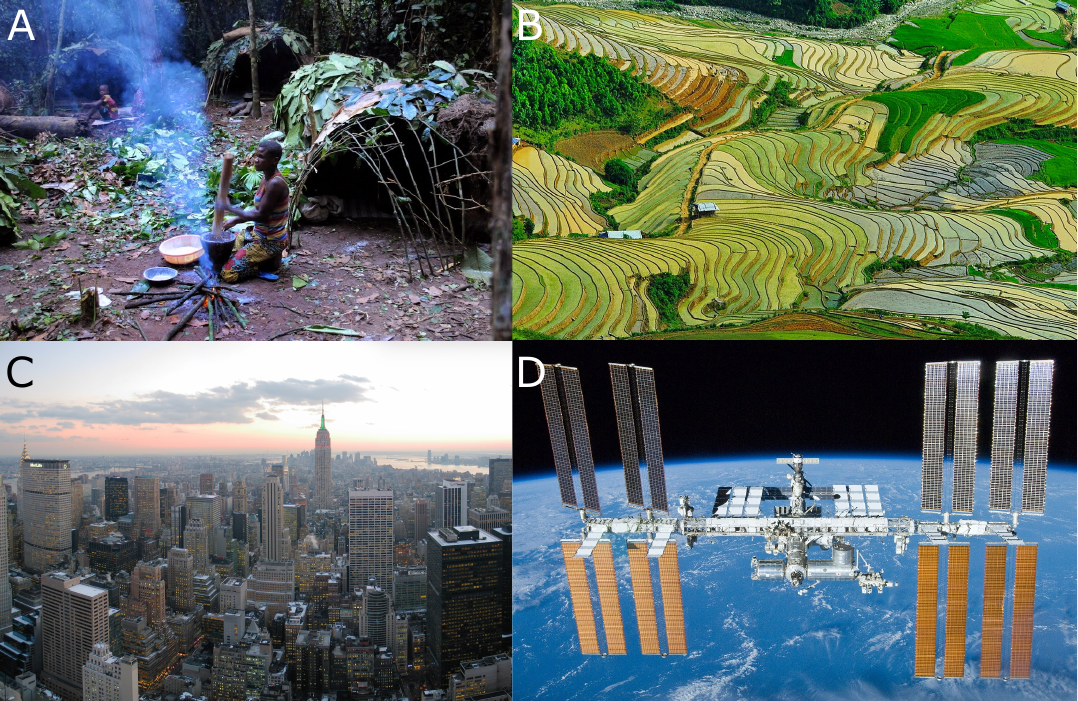}
\caption{\label{fig:Niche diversity} Examples of the niche diversity expressed by the human species in the 21st century: A) A Bayaka hunter-gatherer camp in the Central Africa Republic; B) Rice terraces in the Mu Cang Chai District, Vietnam; C) New York City; D) The international space station. Photo attributions: A) jmgracia100, CC BY-SA 4.0; B) Vu Hung, CC BY-SA 3.0; C) Daniel Schwen, CC BY-SA 2.5; D:) NASA Crew of STS-132, Public domain.}
\end{figure}

Discussions of the human niche in anthropology invariably focus on a subset of ecological traits considered to be unique in comparison to non-human primates \cite{kaplan2000theory, chapais2009primeval, van2016primate, kappeler2010mind}. Such traits might include alloparenting; reciprocal altruism; sharing with non-kin; language; technology; art; belief systems; institutions; non-genetic inheritance; large brains; complex social structures; social learning; and a unique diet \cite{kaplan2000theory, roberts2018defining, fuentes2015integrative, fuentes2016extended, fuentes2017creative, chapais2009primeval, brown1991univ}. While it is important to understand the evolutionary context from which the modern human niche evolved, there is now a wide diversity of lifestyles capable of sustaining human life. This diversity of lifestyles reflects the context-dependent nature of the human niche. The socioeconomic, socioecological, and sociopolitical diversity of these lifestyles demonstrate the array of anthropological solutions to the wide diversity of environments with which humans interact in the 21st century, from foraging to space travel (Figure \ref{fig:Niche diversity}). To explore this diversity we first need to define what we mean by a niche.

\subsection{Classic niche theory}
\begin{figure}[ht!]
\centering
\includegraphics[width=0.75\textwidth]{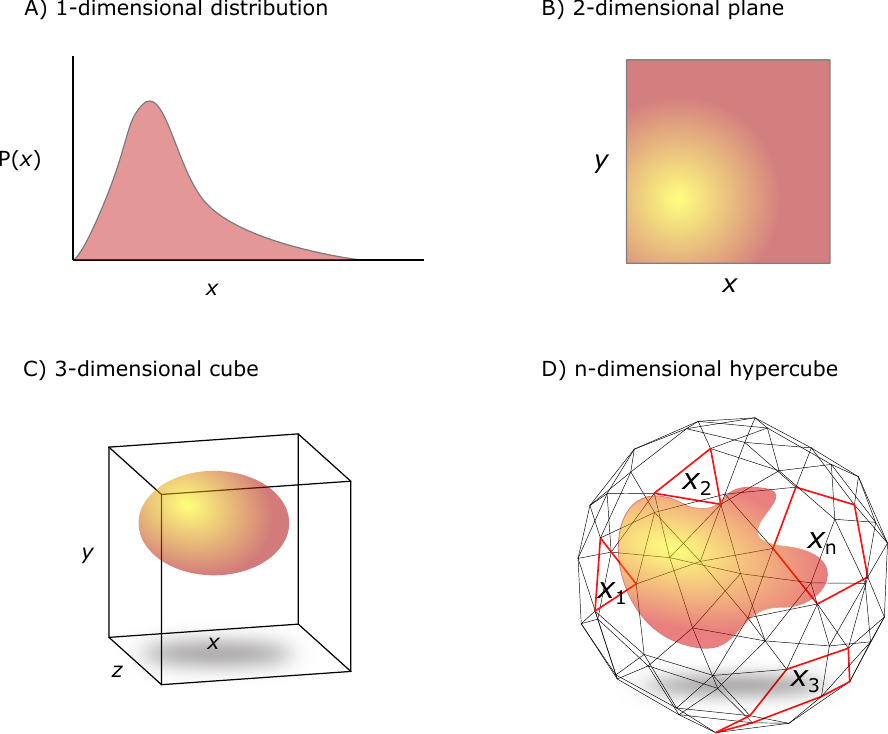}
\caption{\label{fig:n_dimension} The dimensionality of niche space. A) A standard univariate probability distribution of a single variable; B) A bivariate interaction of two variables, such as a scaling plot; C) The interaction of three variables represented as a cube in \textit{x}, \textit{y}, \textit{z} space; D) A conceptual rendering of an \textit{n}-dimensional hypercube defined by \textit{n} faces of \textit{n} interacting variables.}
\end{figure}

The ecological niche was first defined by Joseph Grinnell in 1917 \cite{grinnell1917niche} and has been debated ever since. Although the niche has become a core concept in ecology, much like the term ``culture'' in anthropology, there is no single definition. Indeed, the term niche refers to the abstract operating space of an organism in its environment, but different definitions stress different aspects of this operating space. Famously, Eugene Odum described the niche as an organism's ``profession'' and the habitat as an organism's ``address'' \cite{odum1971fundamentals}. To Grinnell the niche is an attribute of the environment defined by the constraints within which an organism occurs \cite{grinnell1917niche}. As such, the Grinnellian niche is the distributional potential of a species, and this is the usual basis of ecological and eco-cultural niche modeling \cite{peterson2003predicting, banks2006eco}. The Eltonian niche \cite{elton2001animal} is similar in that it is also an attribute of the environment, but emphasizes the ecological role different species may play in an environment while competing for the same niche \cite{colwell2009hutchinson}. The Hutchinsonian niche builds off these ideas but inverts them to emphasize that a niche is an attribute of the organism not the environment, and so adds a crucial evolutionary perspective into the foundational ecological concept \cite{brown1995macroecology}. The Hutchinsonian niche is the environmental space bounded by the abiotic and biotic interactions of an organism that allows positive population growth \cite{hutchinson1957multivariate}. This space is quantified as an \textit{n}-dimensional hypervolume \cite{hutchinson1957multivariate, blonder2018new, holt2009bringing} (see Figure \ref{fig:n_dimension}). Importantly, because the Hutchinsonian niche is an endogenous feature of an organism as opposed to an exogenous feature of the environment, it is a source of natural selection as any heritable change in abiotic or biotic interactions that impacts survival and reproduction leads to population growth, which leads to changes in the genome \cite{brown1995macroecology}. Most anthropological uses of the term niche implicitly use a Hutchinsonian-like definition \cite{fuentes2015integrative, fuentes2016extended}, as discussed by Hardesty in the 1970s \cite{hardesty1975niche, hardesty1977ecological}. 

Hutchinson further identified two types of niche; the \textit{fundamental} niche is determined by the physiological range of tolerances of an organism to its environment independent of all other biotic interactions \cite{soberon2017fundamental}; the \textit{realized} niche is the subset of niche space available to an organism given the environmental availability of those physiological conditions, and the limitations imposed by competition from predators and other species \cite{soberon2017fundamental}. As such, by construction, the fundamental niche of a species is always greater than the realized niche because species distributions are always limited by competition \cite{colwell2009hutchinson, soberon2017fundamental}. The physical mapping of the realized niche into geographic space is essentially Odum's definition of the habitat \cite{odum1971fundamentals, colwell2009hutchinson}. 

However, for a species as geographically distributed and diverse as the human species, differentiating between the fundamental and realized niche is unclear. For example, it could be argued that the fundamental human niche is best defined by the socioecological lifestyles of hominin populations of Middle Pleistocene Africa, 300-200 KYA from which behaviorally-modern humans emerged. It could also be argued that the fundamental human niche is the current global distribution of the human species encompassing all its social, cultural, political, and economic diversity and asymmetry. But, by definition, this distribution must be a realization of a Grinnellian potential. If so, we should also include McMurdo station in Antarctica, the International Space Station, rovers on Mars, and Voyager 1, now $\sim14.6$ billion miles from Earth, in which case the realized human niche is considerably larger than the fundamental human niche. Such definitional confusion highlights a limitation of the classic niche concept when referring to the human species: because the human niche (both realized and fundamental) has been growing, expanding, and diversifying over hundreds of thousands of years the definition of the human niche is largely context-dependent and time-dependent. While classic niche theory may be useful for considering the ecological context of a well-defined human population when both time and space can be effectively held constant \cite{banks2006eco, banks2008human, banks2013human}, if we wish to consider the evolutionary diversification of the human niche, we need more dynamic theory.

\subsection{Ecosystem engineering and niche construction}
Jones et al. \cite{jones1994organisms} defined ecosystem engineering as the ability of organisms to modify their habitats (i.e., their realized niches) by altering abiotic or biotic interactions that impact survival or reproduction. This definition is similar in many respects to the Hutchinsonian niche \cite{hutchinson1957multivariate}. Ecosystem engineers impact resource availability to other species, either positively or negatively \cite{jones1997positive}, not through trophic interactions per se, but through the physical construction of artifacts and the modifications of materials. Importantly, costly physical constructions (such as beaver dams, prairie dog burrows, fish weirs, or skyscrapers) can be inherited conveying non-genetic fitness-related benefits to future generations. Humans are the ultimate ecosystem engineers \cite{smith2007ultimate}, a legacy inherited from our hominin ancestry \cite{faith2021rethinking, braun2021ecosystem, wrangham2009catching}.

More recently, the Hutchinsonian niche and the concept of ecosystem engineering formed the basis of niche construction theory \cite{odling2013niche}, a theoretical convergence of evolutionary biology and theoretical ecology that focuses on population level changes resulting from habitat modification and the fitness consequences of ecological inheritance through the common denominator of ecosystem engineering \cite{barker2014integrating}. As a result, there is much conceptual overlap between the ecological concept of ecosystem engineering and the evolutionary concept of niche construction. However, ecological engineering tends to focus on the fitness impacts of habitat modification at the level of an organism over ecological time scales, whereas niche construction tends to concern population level changes over evolutionary time scales \cite{odling2013niche}.

\subsection{Human niche construction}
The concept of niche construction was quickly adopted into anthropology as it neatly captures both the conceptual complexity of human-environment interactions while referring to a much broader range of ecological behaviors \cite{laland2001cultural, laland2010niche, odling2013niche}. The term niche in anthropology is now most commonly used in the context of niche construction: Google Scholar reports that as of March 2023, $\sim$2,500 papers include the term "human niche construction". However, despite the popularity of the term, the ``human niche'' or it's ``construction'' are rarely defined, operationalized, or quantified. When offered, definitions of the human niche are often conceptual and generic \cite{binford2019constructing, fuentes2015integrative, fuentes2016extended}. Moreover, a well-recognized theme to the literature on human niche construction is the empirical difficulty of operationalizing the term using anthropological data beyond a handful of well-explored narrative case studies \cite{boivin2016ecological, gerbault2011evolution, obrien2012genes, bird2013niche, obrien2021genes}. The lack of definitional clarity may be partly an historical accident because in the foundational text on niche construction--Odling-Smee \textit{et al.} (2003) \cite{odling2013niche}-- the authors include only a limited introduction to the concept of the ecological niche (2 pages out of a 470-page book; pages 37-39) and rely more heavily on the related concept of ecological engineering. 

\section{Measuring dimensions of the human niche}

The central difficulty of operationalizing niche construction in anthropology is that the relevant time-scales fall within an anthropological blind spot. Niche construction addresses habitat modifications and their outcomes at a multi-generational time scale, which is too long to be observed empirically in most ethnographic field studies, and too short to be observed in most archaeological studies. Evolutionary changes over deep time, as observed in the hominin record, for example, may imply the evolutionary interaction of genetic and cultural processes, and behavioral ecology studies of forager behavior may imply the interaction of technology and life history, for example, but neither measure human dynamics at the necessary scale. As such, anthropological data rarely capture the resolution of time-series data required \cite{obrien2021genes}.

If the ultimate goal is to understand how the human niche varies over time and space, a viable alternative to time-series data are comparative data, which are much more common. Here, rather than modeling niche construction directly, the comparative approach allows us to measure variation over different types of human niches at different scales, both within and across species, populations, levels of socioeconomic organization, and time. In this way, human niche diversity can be measured empirically and niche construction can then be inferred and tested. 

\begin{figure}[ht!]
\centering
\includegraphics[width=0.75\textwidth]{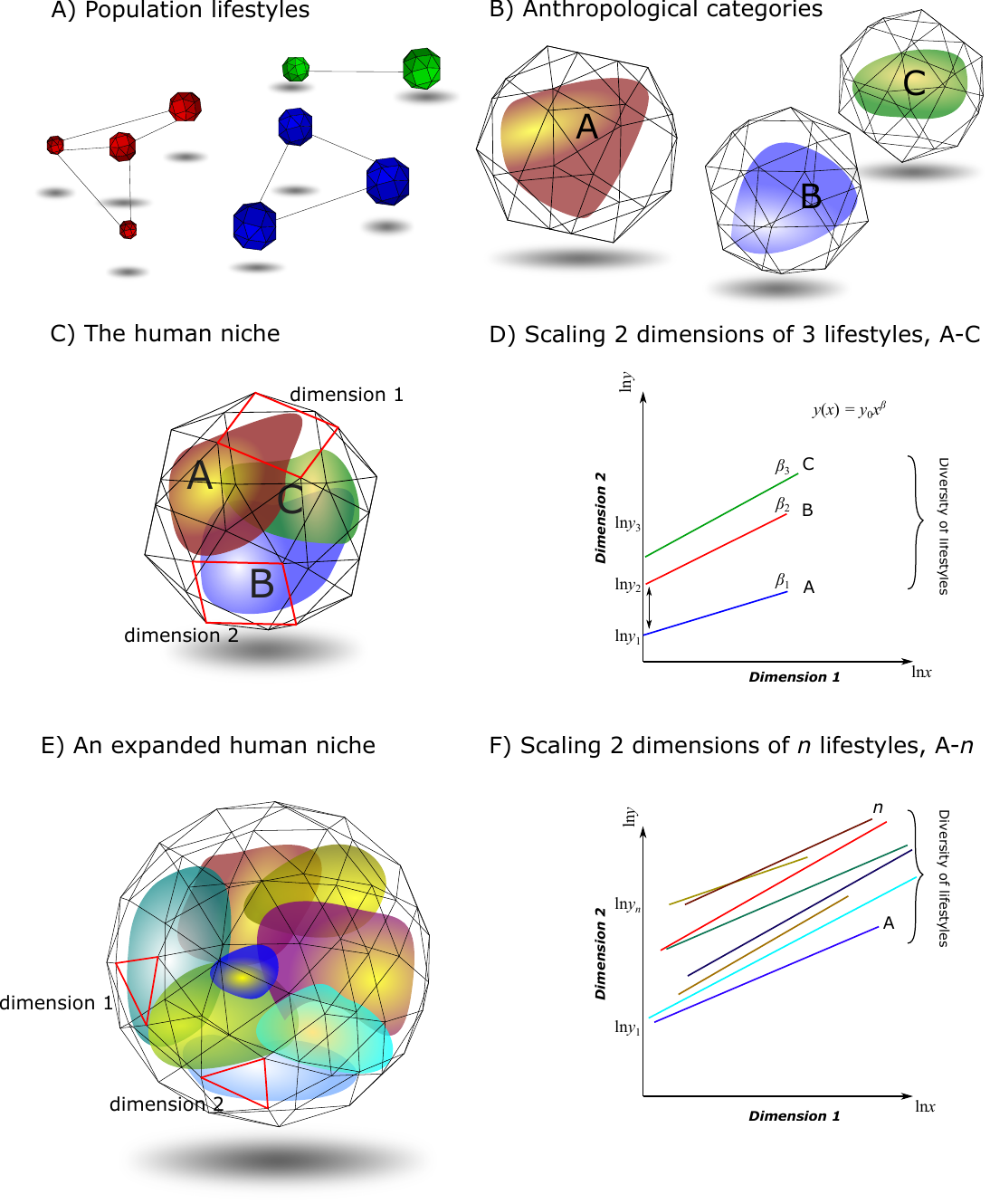}
\caption{\label{fig:Lifestyle} The dimensionality and scales of human niche space. A) Discrete populations have their own niches but share lifestyles; B) Clusters of lifestyles form anthropologically relevant clusters; C) The human niche is the combination of clusters, lifestyles, and populations; D) A log-log plot of the bivariate scaling of 2 dimensions on the available \textit{n}-dimensions for 3 lifestyles. E) An example of an expanded human niche with a greater diversity of lifestyles, a larger overall scale, and more dimensions along which to vary. F) A log-log plot of the bivariate scaling of 2 dimensions on the available \textit{n}-dimensions for \textit{n} lifestyles. Note that in comparison to panel D there is a greater diversity of lifestyles and expansion along both the \textit{x}  the \textit{y} dimensions.}
\end{figure}

\subsection{Defining the human niche}
The ecological niche is a characteristic of a population composed of individuals who vary in their adaptive strategies within some range of variation defined by \textit{n} probability distributions. Variation across human populations occurs at multiple scales of anthropological relevance. For example, holding the human genome, metabolism, and cognition constant, there are meaningful qualitative and quantitative differences in the lifestyles of human hunter-gatherer populations in comparison to say industrialized market-based national populations. Populations from both lifestyles may be measured meaningfully along similar axes of variation (population size, density, energy consumption, productivity etc.), but the fundamental scales at which those measurements are made differ qualitatively in many ways. 

Here I define the human niche hierarchically in a way that can be operationalized using models from scaling theory and data (Figure \ref{fig:Lifestyle}): \textit{The human niche} refers to the \textit{n}-dimensional hypervolume occupied by the human species at any one time in the past, present, or future. The human niche is comprised of a diversity of populations, each of whom have a \textit{lifestyle} defined by the interaction of demographic, economic, ecological, biogeographic, technological, sociopolitical, historical and cultural evolutionary traits, all of which are random variables representing a range of variation along a single dimension of the \textit{n}-dimensional human niche. Lifestyle traits are shared among populations through autocorrelation in space, phylogenetic history, or convergence. Shared lifestyles that organically cluster may form analytically-meaningful coarse-grained categories that capture some larger aspect of relevant anthropological categorization based on the question at hand (i.e., forager/farmer/etc., egalitarian/non-egalitarian, or mobile/sedentary/equestrian/etc.).
In the scaling framework, these functions are operationalized by relating dependent and independent variables as simple scaling relationships defined by power functions of the form

\begin{equation}
    y(x) = y_0x^{\beta}
    \label{eq:scaling}
\end{equation}

\noindent{}where \textit{y} is some dependent variable of interest, such as as measure of life history or economic productivity, $y_0$ is a normalization constant, \textit{x} is an independent variable, such as population size or energy consumption, and $\beta$ is a scaling exponent, or elasticity, describing the proportional change in \textit{y} to a proportional change in \textit{x}. Note that power functions of the form of equation \ref{eq:scaling} are flexible as they allow both linear ($\beta=1$) and nonlinear ($\beta \neq 1$) responses. Equation \ref{eq:scaling} can then be extended to incorporate multi-level factors (i.e., relevant anthropological categorization) and additional covariates as required. We can then estimate the parameters of equation \ref{eq:scaling} by taking the logarithm of both sides and fitting the equivalent linear model to available data. Having defined and operationalized the niche concept, we can now consider a number of fundamental dimensions of the human niche relevant at all anthropological scales: metabolism; sociality; cognition; and computation.

\subsection{The metabolic dimension}
Body size is of fundamental importance in ecology as it determines the metabolic demand of an organism, and the resulting energy trade-offs shape all aspects of a species evolved life history, physiology, much of its behavior \cite{brown1995macroecology, brown2004toward}. Body size also influences the relative complexity of the environment as differently-sized organisms will encounter resources on the same landscapes at different rates \cite{haskell2002fractal, ritchie2009scale}.  Body size thus sets the scale of the niche in mammals \cite{wilson1975adequacy, woodward2005body}. 

\begin{figure}[ht!]
\centering
\includegraphics[width=1\textwidth]{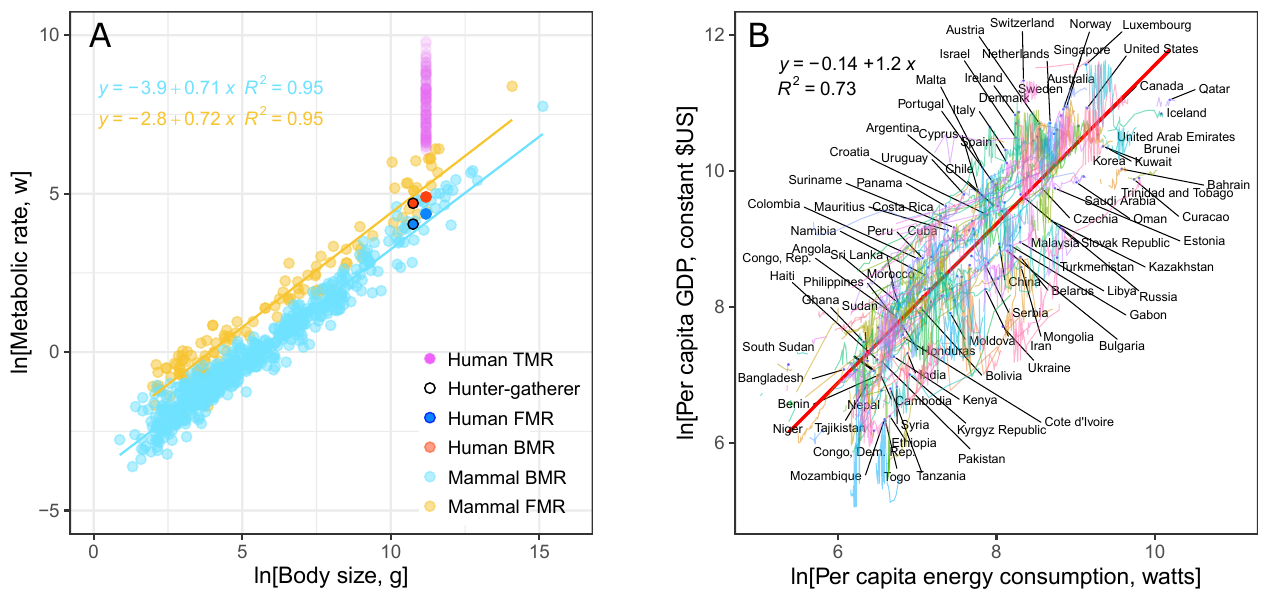}
\caption{\label{fig:Metabolisms_nations} Scaling of energy consumption, body size, and productivity across mammals and countries. A) Metabolic rates as a function of body size across humans and mammals. Both basal metabolic rates (BMR) and field metabolic rates (FMR) scale similarly with body size across mammals, and for hunter-gatherers and industrial humans. The major difference in industrial human energy budgets is the total metabolic rate (TMR), or extrasomatic energy budget (purple points), which in the most consumptive countries can be over 300-times the expected field metabolic rate of a human-sized mammal. Industrial lifestyles have had little effect on the human somatic metabolic budget. In figure \ref{fig:Metabolisms_nations}A, hunter-gatherer data come from field studies of the Hadza in Tanzania \cite{pontzer2015energy}, who have smaller average body sizes than the average body sizes of the sample of industrial humans. B) Gross domestic product as a function of energy consumption across countries for the period 1960-2021 using data from the World Bank. Each line is the time series of a country. }
\end{figure}

For an animal of a given body size, variation (or diversity) within the niche is then primarily dependent on environmental variation. For humans, niche diversity is also dependent on technology, economy, and demographic scale (i.e., population size). 

Empirically, across mammals the size-dependent basal metabolic rate goes as
\begin{equation}
    B_{BMR}(M) = b_0M^{3/4}
\end{equation}

and, similarly, the size-dependent field metabolic rate goes as
\begin{equation}
    B_{FMR}(M) = b_1M^{3/4}
\end{equation}
as shown in Figure \ref{fig:Metabolisms_nations}. Empirically, on average $b_1$ is about three times greater than $b_0$ and so the basal metabolic rate is approximately $1/3$rd of the field metabolic rate across mammals, sometime termed the sustained metabolic scope \cite{peterson1990sustained}, or the physical activity level \cite{pontzer2015energy}. Figure \ref{fig:Metabolisms_nations} shows that for humans (both hunter-gatherer and industrial) both the basal and field metabolic rates are much as expected for a mammal of the equivalent body size though the sustained metabolic scope (FMR/BMR) of industrial humans is slightly less than in hunter-gatherers; 1.93 vs. 1.72. Therefore, industrialization has had little effect on the somatic energy budget of humans beyond a slight decrease in physical activity levels. However, Figure \ref{fig:Metabolisms_nations} shows that the dramatic difference in the industrial human energy budget is what can be termed the extrasomatic metabolic rate, or the total energy consumed per capita. The average extrasomatic metabolic rate across countries (using 2014 data from the World Bank) is about 3,600 watts ranging from a minimum of 200 watts in South Sudan to a maximum of 26,000 watts in Qatar. In the US the average extrasomatic metabolic rate in 2014 was 9,500 watts. Therefore, in developing countries, such as South Sudan, about 40\% of the total daily energy budget is used to support the basal metabolic function of an individual, and the extrasomatic metabolic rate is not much more than the field metabolic rate. In the wealthiest countries, the equivalent percentage is $\sim0.003\%$. That is to say, the average person in the most energetically consumptive countries consumes about 325-times more energy than their basal metabolic requirements, or over 100-times more energy than the average person in the least consumptive countries. 

Figure \ref{fig:Metabolisms_nations}B  zooms into the pink line of data on figure \ref{fig:Metabolisms_nations}A to consider the productivity of energy consumption at a national level, measured here as Gross Domestic Product (GDP) over a 60-year time series. Not only is there a 100-fold difference in the energy consumption of countries, there is also a nearly 500-fold difference in per capita wealth across countries, from Burundi with an average per capita GDP of $\$384$ to Monaco at $\$160,000$ (2014 dollars). In Brown et al. \cite{brown2011energetic}, we presented similar data on reversed axes to demonstrate the economy of scale in international development, where improvements in well-being (and associated decreases in reproductive rates \cite{burger2011industrial}) necessarily require increased energy consumption, but at diminishing rates. Figure \ref{fig:Metabolisms_nations}B explains why this is the case: per capita GDP increases superlinearly with energy consumption, and so each new watt consumed produces an additional $20\%$ of GDP and associated wealth, well-being, and security. Poor countries in the lower left of Figure \ref{fig:Metabolisms_nations}B aim to follow this developmental trajectory toward the wealth generating economies of countries further along the x-axis by consuming more energy in order to benefit from these increasing returns to scale. At the same time, the average person in those countries to the upper right of the panel need to consume energy at two orders of magnitude the rate of the poorest countries in order to maintain their lifestyles. Effectively, this superlinearity quantifies the Red Queen-like economics of cumulative culture and human niche construction \cite{van1973new, tomasello1993cultural, mesoudi2018cumulative}.

\subsection{The cognitive dimension}

The cognitive niche can be defined as the ability to use knowledge to solve ecological problems, such as overcoming the defense mechanisms of plants or engineering traps to snare prey animals \cite{pinker2010cognitive}. The sociocognitive niche is an extension of the cognitive niche that emphasizes the importance of cooperation, language, theory of mind, and shared intentionality among groups of non-related humans \cite{whiten2012human, tomasello2005understanding}.

\begin{figure}[ht!]
\centering
\includegraphics[width=1\textwidth]{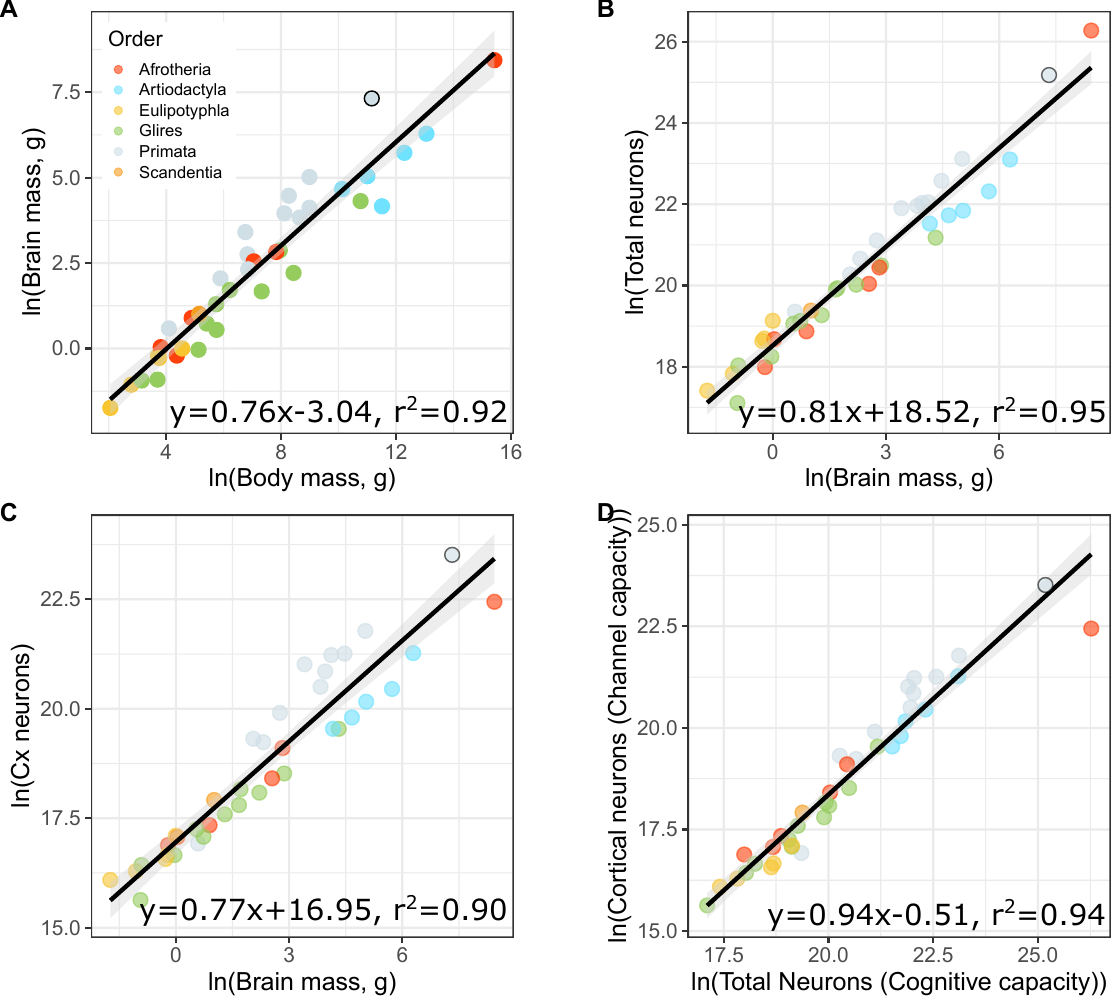}
\caption{\label{fig:brain_order}Scaling brain size and neural anatomy across mammal orders using data from Herculano-Houzel et al. \cite{herculano2015mammalian}. A) Brain mass as a function of body mass. B) The number of cortical neurons as a function of brain mass. C) The number of cortical neurons as a function of body mass. D) The number of cortical neurons as a function of total neurons in the brain. Assuming computational capacity is correlated with cortical neurons \cite{herculano2017numbers}, and channel capacity with the number of neurons in the brain \cite{el2011network, stone2018principles} then computation is (approximately) proportional to the brain's bandwidth across mammals. These data show that the human brain is notable in the class Mammalia, not due to its size per se, but in the number of cortical neurons in the brain. However, the number of cortical neurons in the human brain is consistent with the scaling of total neurons in the brain within mammals.}
\end{figure}

It has long been recognized that mammalian brain size, $S$, scales with body size as $S\sim M^{3/4}$ \cite{martin1981relative, burger2019allometry} (see Figure \ref{fig:brain_order}). It has been proposed for just as long that deviations from this scaling, commonly known as the encephalization quotient, reflect some measure of mammalian intelligence and this is reflected in the observation that the human brain is about 7-times the size of a mammal of equivalent body size \cite{jerison2012evolution, harvey1991comparative}. However, much of this reasoning was largely speculative \cite{deaner2007overall}. Recent data produced by Herculano-Houzel and colleagues \cite{herculano2015mammalian} can be used to quantify the scaling of mammalian neural anatomy. Given that the number of neurons in the brain are biological correlates of cognitive capability \cite{herculano2017numbers} Herculano-Houzel and colleagues argue the computational capacity of the human brain stems not from it's relative size, but from the sheer number of neurons in the human brain ($\sim 86$ billion \cite{azevedo2009equal}) and their particular concentration of neurons in the cerebral cortex \cite{herculano2012remarkable, passingham2021understanding}. For example, while Figure \ref{fig:brain_order}A shows that primates have larger brain sizes for their body mass than most other mammals, \ref{fig:brain_order}B shows that the scaling of the total number of neurons in the brain is remarkably consistent across mammals. Note African elephants (\textit{Loxodonta africana}) have larger brains with more neurons than humans. However, Figures \ref{fig:brain_order}B and C shows that while the scaling of cortical and total neurons with brain size are similar, primates have more cortical neurons for their brain size than other mammals. Indeed, Figure \ref{fig:brain_order}C shows humans have the most cortical neurons than any other species in the data set, largely because humans are large-bodied primates. 

Using these results, we can then speculate--albeit quantitatively--about the nature of computation in the human brain with respect to primates and other mammals (also see \cite{hamilton2022collective}). If we assume the number of cortical neurons is proportional to cognitive capacity of the brain \cite{herculano2017numbers} and that the total number of neurons in the brain is proportional to the brain's channel capacity \cite{el2011network, stone2018principles} (i.e., the informational bandwidth of the brain), then Figure \ref{fig:brain_order}D captures a fundamental feature of computational capacity in mammalian brains. Mammalian cognitive capacity scales linearly with channel capacity, which suggests brains with more neurons are simply capable of more higher level processes because they can handle more information. Human cognitive capacity is a function of the absolute number of neurons available, not due to any particular reallocation of neurons to the cerebral cortex. It then follows that human cognitive capacity is indeed a function of the absolute size of the human brain, but only insofar as humans are large-bodied primates, an order evolutionarily invested in large brains and complex cognitive function \cite{passingham2008special, passingham2021understanding, dunbar1992neocortex, dunbar1998social}. However, the consequences of this cognitive evolutionary inheritance for the diversification of the human ecological niche are considerable.

\subsection{The social dimension}
A crucial consequence of human-like cognition is the ability to cooperate with non-kin through reciprocal altruism \cite{trivers1971evolution}. As such, the human cognitive niche is often termed the \textit{socio}cognitive niche \cite{whiten2012human, tomasello2005understanding}, as the cognitive capacity of a single brain contributes to a much larger distributed social network of similar brains \cite{hamilton2022collective}. This ability underwrites all aspects of human evolutionary biology, ecology, and social organization, from the evolution of the human life history \cite{kaplan2000theory} and family formation \cite{lancaster2017watershed}, to the development of large social networks \cite{chapais2009primeval} and associated economies of scale \cite{hamilton2007nonlinear}. These traits form a complex, robust, inventive, and flexible human socioecology, which allowed hunter-gatherer lifestyles to expand the human biogeographic range across most of the planet in the late Pleistocene, diversifying over the Holocene with the emergence of agricultural, urban, and state economies and institutions. Indeed, the cumulative nature of human cultural evolution characteristic of the Holocene emerged from a long evolutionary history of economic diversity and technological innovation \cite{shipton2015before, tennie2009ratcheting}.

\begin{figure}[ht!]
\centering
\includegraphics[width=1\textwidth]{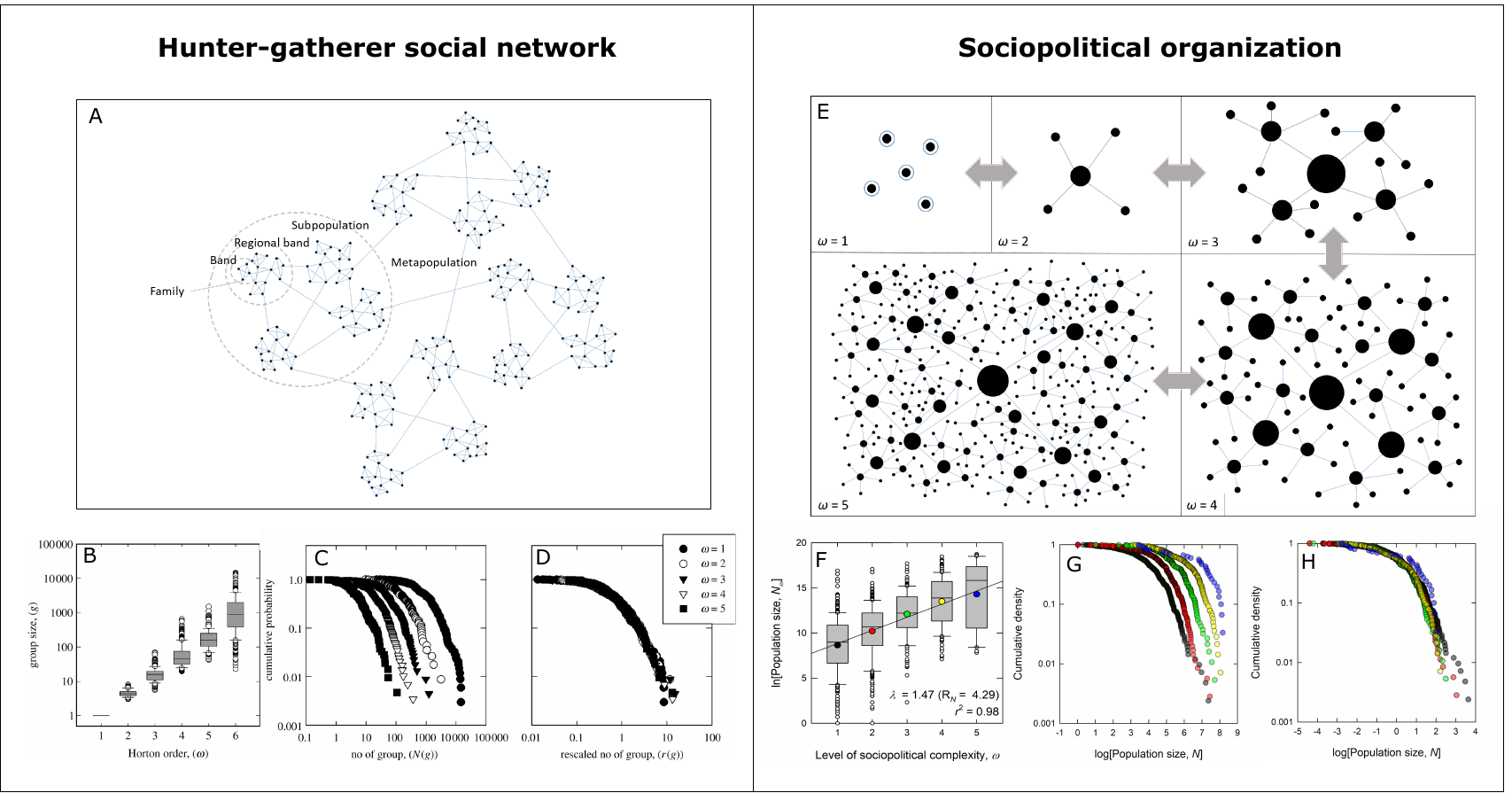}
\caption{\label{fig:Social_comb}The nested, hierarchical, self-similar small-world-like properties of hunter-gatherer metapopulations on the left (panels A-D) and across levels of sociopolitical complexity in traditional societies (panels E-H) \cite{hamilton2007complex}.}
\end{figure}

The adaptability, flexibility, and robustness of human social organization comes from the complex structure of human social networks. Human sociality is fundamentally modular (families; groups; tribes; firms; nations; football teams), nested (small groups or modules are often part of larger networks which are part of larger networks and so on), and fluid (modules fission and fuse at all scales through relocation, employment, marriage, competition, contracts etc.). As such, human social networks are dynamic small-worlds where dense local interactions are connected by much sparser, but vitally important weaker ties \cite{granovetter1973strength}, that serve to bind complex social systems at multiple levels of organization. 

Moreover, the modular, nested structure of human social networks is often statistically self-similar (see Figure \ref{fig:Social_comb}) \cite{zhou2005discrete, fuchs2014fractal, hamilton2007complex, hamilton2020scaling}. That is to say, if we define $\langle g(\omega) \rangle$ as the average group size of some module at the $\omega th$ level of the social network, the branching ratio $B = \langle g(\omega +1)\rangle/ \langle g(\omega)\rangle$ between all adjacent levels is approximately constant, and so the system is statistically self-similar. In hunter-gatherer metapopulations, we find $B\approx4$, and so there are approximately 4 individuals in a family, 4 families in a co-residential group, 4 co-residential groups in a regional group, 4 regional groups in a sub-population, a 4 sub-populations in a metapopulation (\ref{fig:Social_comb}A-B) \cite{hamilton2007complex}. Interestingly, we find a similar fractal-like structure to the hierarchical levels of sociopolitical complexity across traditional human societies, where additional layers of sociopolitical hierarchy are associated with populations that are on average 4-times the size (Figure \ref{fig:Social_comb}E-F) (\cite{hamilton2020scaling}. The robustness of this self-similarity is shown in Figures \ref{fig:Social_comb}C-D, and G-H, respectively, where the entire distributions of groups sizes at all $\Omega$ levels in both data sets collapse onto single scaling functions.

\subsection{The collective computational dimension}
Extensive fitness-enhancing exchanges of energy and information among non-kin through complex social networks facilitates collective computation \cite{hamilton2022collective}. Here, 
collective computation is defined as the ability of complex adaptive systems--such as networked populations of humans--to produce adaptive solutions to problems by inventing, innovating, aggregating, and deploying knowledge across scales \cite{brush2018conflicts}. In effect, social networks facilitate the transmission, filtering, and recombination of ideas originating at the individual level by scaling-up and boosting ideas to impact the ecology and economics of groups. But importantly, the scaling up of solutions is often a symmetry-breaking process \cite{anderson1972more}, resulting in nonlinear returns to scale \cite{west2018scale}. For example, following from equations 2 and 3, within humans we can assume body size is effectively constant, or, more precisely, we can define an average body size with an associated metabolic rate

\begin{equation}
    \langle B(M) \rangle = \langle b_0M^{3/4} \rangle = 1/N \sum_{i=1}^{N} (b_0M^{\beta}).
\end{equation}

We can then scale up the expected metabolic demand of an average individual to the expected metabolic demand of a population of \textit{N} individuals symmetrically
\begin{equation}
    B_N(M)=\langle B(M) \rangle N
\end{equation}
and so the expected metabolic demand of a population is simply the individual demand multiplied by the population size. If we then consider a hunter-gatherer population where dietary resources must be foraged in space, we define the home range of an individual as

\begin{equation}
    H(R)= \frac{\langle B(M) \rangle} {R}
\end{equation}
where $R$ is a resource supply rate measured in units of energy per unit time, such as net primary productivity (g C $m^{-2}$ $yr^{-1}$): $R$ is dependent on many interacting biotic and abiotic factors, such as biodiversity, temperature, precipitation, and spatial structure \cite{brown2014there, hamilton2020diversity, ritchie2009scale, brown2004toward}. The total area $A$ used by $N$ individuals to meet their combined metabolic demand $B_N(M)$ per year is

\begin{equation}
    A(N)= H(R)N= \frac{\langle B(M) \rangle} {R} N= h_0 N^{\alpha}
\end{equation}
where $h_0$ is a normalization constant and $\alpha$ is a scaling exponent. Empirically, $\alpha<1$ in many small-scale human subsistence populations \cite{freeman2016socioecology, walker2014remote, hamilton2012human, hamilton2007nonlinear} indicating economies of scale in the energetics of human spatial ecology as the area per individual decrease with population size; $A(N)/N \propto N^{\alpha-1}$. This spatial sublinearity is observed in urban populations where larger cities are also denser \cite{bettencourt2007growth, burger2017extra} and we also observe similar sublinearity in the scaling of total energy use and population size across countries \cite{hamilton2012human}. In these examples, the ecological symmetry of resource availability in space is broken (presumably) by cooperation among individuals through reciprocal altruism and information sharing. That is to say, the ability of an individual to meet their own energy demand is not solely a property of the individual, but an outcome of the larger collective at all levels of human social organization in all niches, from foraging populations to industrialized market economies. The collective computation here then is the deviation in linearity resulting in an economy of scale in space use that emerges from the collective behavior of the larger network unachievable by individuals working alone.

\section{Discussion}
The human niche is defined as the operating space of human populations as they interact with their biotic, abiotic, and social environments. This is a multidimensional space, an \textit{n}-dimensional hypervolume the dimensions of which are determined by the diversity of lifestyles employed to meet the combined resource demands of health, reproduction, and well-being in differing circumstances across the planet (and beyond). The complexity of defining the human niche stems from the ways in which human-environment interactions have evolved over time and space. While the fundamental scales of human life history and environmental interactions are set by the body size of the human organism, the evolutionary expansion of the human niche is determined by the ways in which technologies have evolved over time to reduce uncertainty or to increase the productivity of these interactions. The diversity of the human niche at any one time in the past is thus a function of both the biogeographic distribution of human populations, the historical availability of technologies to those populations, and the energy flux those technologies generate.

\begin{figure}[ht!]
\centering
\includegraphics[width=1\textwidth]{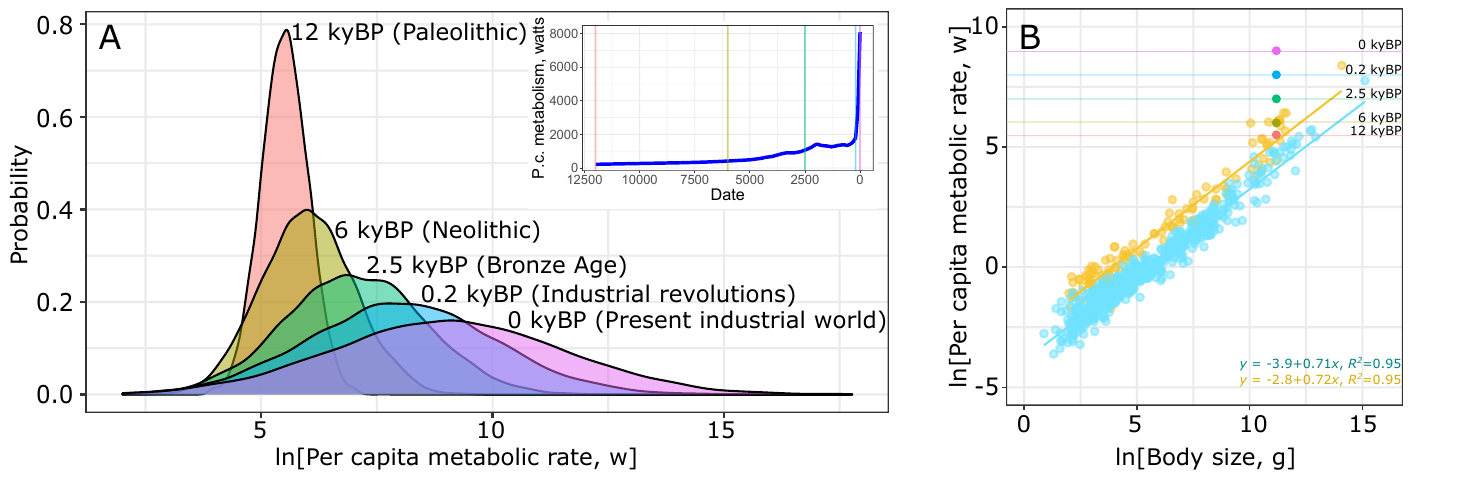}
\caption{\label{fig:Metabolic_dists}The lifestyle diversification, niche expansion, and metabolic evolution of the human species. A) A schematic of the evolution of metabolic scope of the human species based on data from Morris (2013) \cite{morris2013measure}. From the end of the Paleolithic per capita metabolic consumption has both increased and diversified. In the industrial world of the present, per capita metabolism is both at it greatest scale and scope. Inset is a figure of the growth of per capita metabolism over the last 12,000 years, using the same data source. B) The scaling of metabolic rates and body size across mammals highlighting the increased scale of human per capita metabolic consumption over the last 12,000 years. C) The expansion and divesification of the human niche and lifestyles. Each lifestyle is comprised of several populations, and the niche at any one time, \textit{t}, is comprised of several lifestyles. After some time, \textit{dt}, the human niche has expanded to include a wide diversity of lifestyles.}
\end{figure}

Figure \ref{fig:Metabolic_dists} uses data from Morris (2013) \cite{morris2013measure} to visualize the evolution of human metabolic diversity over the past 12,000 years. The metabolic dimension of the human niche has evolved over time from a Paleolithic world of hunter-gatherers to encompass a broad diversity of lifestyles as new technologies, institutions, and economies evolved over the Holocene, supporting larger, denser, and more consumptive populations. In 2006, the human species became an urban species and the majority of humans now live in cities \cite{bettencourt2007growth, bettencourt2010unified}. The environmental impacts of this transition have led to the newly defined geologic era of the Anthropocene \cite{crutzen2006anthropocene}. As new lifestyles emerged over the Holocene, many, but not all ancestral lifestyles were  displaced, and so many subsistence-scale lifestyles, including a few hunter-gatherer societies, still occur as part of human niche diversity throughout much of the world. That is to say, the generation of new niches has outpaced the extinction of ancestral niches, and so more is different \cite{anderson1972more}. As such, today, the metabolic and technological scope of human niche diversity is at its broadest in human evolutionary history.

The majority of the technological and metabolic expansion of the human niche has occurred over the last few hundred years, since the onset of the first industrial revolution in 1760 (Figure \ref{fig:Metabolisms_nations}A inset), fueled by increased access to fossilized biomass stored below the Earth's surface \cite{smil2018energy, brown2014macroecology}. If the human population before the development of agriculture was on the order of 10 million or so, the last 12,000 years has seen a 800-fold increase in human biomass, which reached 8 billion in November, 2022. Coincident with this scale of population growth, the total somatic metabolic demand of human biomass in a preagricultural world would have been on the order of a billion watts (10 million x 100 w), whereas today, the total (somatic + extrasomatic) metabolic consumption of human biomass and its supporting infrastructure is on the order of 20 trillion watts (2014 data of total energy consumption across nations). The transition from the hunter-gatherer world of the late Pleistocene through the expansion of the agricultural world of the Holocene and into the urban world of the Anthropocene required a 17,000-fold increase in the total energy flux of the human species channeled through increasingly complex economies of scale, institutions, and technologies. 

The enormous scale and rapidity of this growth is well-described by scaling behavior, such as shown in Figure \ref{fig:Metabolisms_nations}B. Similar superlinear returns to scale are now well-documented in the economic productivity of cities, both in the present \cite{bettencourt2007growth, bettencourt2010unified, bettencourt2013origins} and in the past \cite{ortman2014pre, ortman2015settlement, lobo2020urban}, suggesting increased returns to economic productivity may be general features of human sociality over deep evolutionary history. This hypothesis is supported by well-documented economies of scale in energy consumption and space use in hunter-gatherer populations \cite{hamilton2007nonlinear, hamilton2018scaling, lobo2022scaling, haas2015settlement}.

\begin{figure}[ht!]
\centering
\includegraphics[width=1\textwidth]{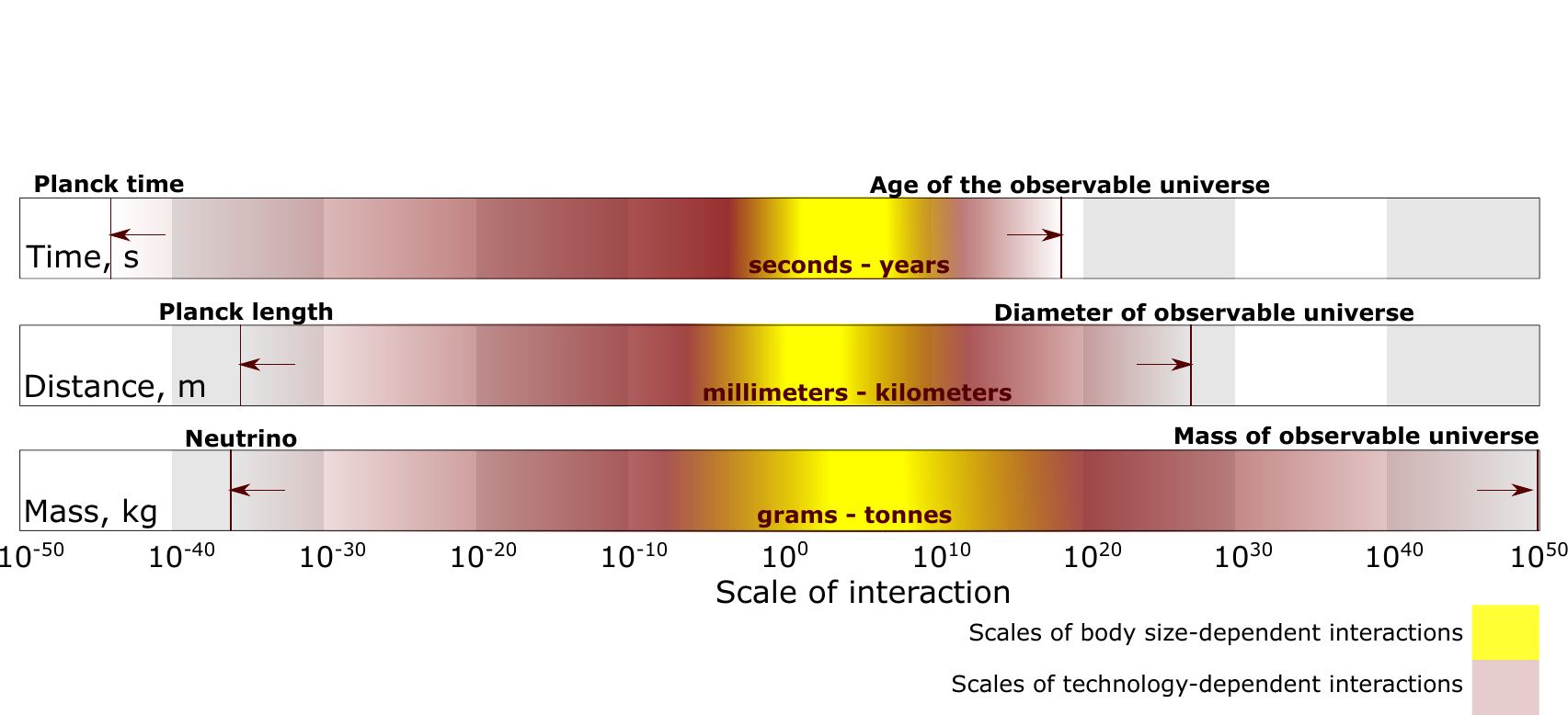}
\caption{\label{fig:Scales of interaction}Changing scales of human interactions with time, space, and matter that define the scope of the human niche. Typical time scales most relevant to the human experience range from seconds to years; scales of space range from millimeters to kilometers; and scales of matter range from grams to tonnes. These scales of daily experience are determined primarily by body size. Recent advances in science, technology, and engineering have maximized the scales from which we can now recover information from the universe, from the smallest Planck scales to the size of the observable universe.}
\end{figure}

Figure \ref{fig:Scales of interaction} illustrates the expanded scales of knowledge facilitated by human niche expansion over the past few hundred years. The scales at which humans interact with the world are limited to a few orders of magnitude, set primarily through body size \cite{ritchie2009scale} and correlate with the scales at which inferences are made by traditional societies. For the vast majority of hominin evolutionary history relevant scales of time range from seconds to years, distance from millimeters to kilometers, and mass from grams to tonnes. While cultural evolutionary processes have manipulated phenomena at much smaller scales in the past, such as the genetic engineering of biodiversity during the development of domesticated plants and animals and the evolution of farming ecologies, cultural evolution does not require an in-depth understanding of the causal mechanisms involved \cite{derex2019causal}. However, with the birth of science and mathematics, the relevant scales at which humans are capable of manipulating matter and developing predictive theories have expanded to both the smallest and largest conceivable scales. Of course, the extent to which our understanding is set by limits of our cognition are open to debate \cite{wolpertArxiv}.

Nevertheless, expanded scales of inference have led to the development of transformative technologies in recent decades such as cell phones, global positioning systems, and nuclear fission, as well as the next generation of potentially-transformative technologies such as artificial intelligence \cite{bostrom2017superintelligence, mitchell2019artificial, tegmark2018life}, quantum computing \cite{preskill2018quantum}, and nuclear fusion \cite{doe2023}. The impact of such technologies on the expanding human niche in the future are significant yet unpredictable. For example, access to effectively unlimited energy through nuclear fusion would lead to unprecedented scales of interaction with earth systems and beyond. Similarly, widespread, cheap and easy access to AIs and quantum computation will lead to unprecedented access to information processing. The potential interaction of these two dimensions is the source of much debate \cite{bostrom2017superintelligence, kurzweil2014singularity}. As the human niche expands along dimensions of computation and energy acquisition it is tempting to draw an analogy with the expansion of the global economy since the industrial revolution. Humans can at best hope to influence the global economy locally, but do not control it globally, and so most of the human species is subject to the fluctuations of markets and national economies seemingly well beyond the reach of any predictive theory.

Perhaps the most ambitious proposal for the intentional diversification of human lifestyles is the expansion of the human niche beyond Earth and into the solar system \cite{siddiqi2018beyond}. While humans have been actively exploring space since 1961, the duration of longest human spaceflight thus far is 487 days by Soviet cosmonaut Valeri Polyakov on the Mir space station in the 1990s. The International Space Station has been active for 24 years and continuously occupied for over 22 years. Therefore, while human technology can facilitate temporary exploration beyond Earth's atmosphere, including deep space (Voyager 1 is 45 years old and now in interstellar space beyond the Solar system $\sim$15 billion miles from Earth), humans have not yet reproduced in space and so space would not be included in most definitions of the human ecological niche. Clearly, discussions of space colonization, including multi-generational deep space missions, and establishing permanent lunar bases and the colonization of Mars would require reproduction with the express intent of expanding the human niche to include extraterrestrial environments
\cite{white2001humans, levchenko2021mars, goldsmith2022end}. While such ideas break no laws of physics, the physiological, psychological, logistical, political, and economic stresses of maintaining and growing viable human populations over the long term in such hostile environments are immense \cite{goldsmith2022end, afshinnekoo2020fundamental, gunga2020human, cahill2020space, wheeler2001plant, kanas2008space, durante2011physical}. Irrespective, scaling theory tells us that at a minimum, reproductively and socially viable human societies on Earth are comprised of modular, nested, yet fluid metapopulations at all scales, maintained by both economies of scale in resource consumption, and increasing returns to scale in productivity. While on Earth these universal features of human populations self-organize, emerging from local optimizations played out at multiple scales and dimensions over time and space \cite{hamilton2022collective}, how self-organization could be engineered into a process as strictly controlled as space travel poses even greater challenges.


\subsection{Concluding thoughts}
Perhaps the most fundamental insights offered by scaling theory to anthropology are ontological and epistemological. The ontological commitments of scaling theory recognize that the anthropological world we seek to describe is complex, hierarchical, and scale-dependent. The epistemology of scaling theory embodies these commitments by requiring that the models we make of the world are similarly complex, hierarchical and scale-dependent. But because these scales are nested and weakly emergent, there must be a universal core set of mechanisms conserved across the major transitions that demarcate human diversity. In this sense, a successful effective scaling theory of anthropology would describe both this universality and emergence (or both the symmetry and its breaking). Thus, while scaling theory offers a fresh perspective to anthropology, it also poses a major conceptual challenge to developing an anthropology of complex adaptive systems.

\bibliographystyle{unsrt}
\bibliography{main}

\begin{thebibliography}{100}

\bibitem{brown1995macroecology}
James~H Brown et~al.
\newblock {\em Macroecology}.
\newblock University of Chicago Press, 1995.

\bibitem{binford1962archaeology}
Lewis~R Binford.
\newblock Archaeology as anthropology.
\newblock {\em American antiquity}, 28(2):217--225, 1962.

\bibitem{white2016evolution}
Leslie~A White.
\newblock {\em The evolution of culture}.
\newblock Routledge, 1959.

\bibitem{binford2019constructing}
Lewis~R Binford.
\newblock {\em Constructing frames of reference: an analytical method for
  archaeological theory building using ethnographic and environmental data
  sets}.
\newblock University of California Press, 2019.

\bibitem{kaplan2000theory}
Hillard Kaplan, Kim Hill, Jane Lancaster, and A~Magdalena Hurtado.
\newblock A theory of human life history evolution: Diet, intelligence, and
  longevity.
\newblock {\em Evolutionary Anthropology: Issues, News, and Reviews: Issues,
  News, and Reviews}, 9(4):156--185, 2000.

\bibitem{chapais2009primeval}
Bernard Chapais.
\newblock {\em Primeval kinship: How pair-bonding gave birth to human society}.
\newblock Harvard University Press, 2009.

\bibitem{van2016primate}
Carel~P Van~Schaik.
\newblock {\em The primate origins of human nature}.
\newblock John Wiley \& Sons, 2016.

\bibitem{kappeler2010mind}
Peter~M Kappeler and Joan~B Silk.
\newblock {\em Mind the gap}.
\newblock Springer, 2010.

\bibitem{roberts2018defining}
Patrick Roberts and Brian~A Stewart.
\newblock Defining the ‘generalist specialist’niche for pleistocene homo
  sapiens.
\newblock {\em Nature Human Behaviour}, 2(8):542--550, 2018.

\bibitem{fuentes2015integrative}
Agust{\'\i}n Fuentes.
\newblock Integrative anthropology and the human niche: toward a contemporary
  approach to human evolution.
\newblock {\em American Anthropologist}, 117(2):302--315, 2015.

\bibitem{fuentes2016extended}
Agust{\'\i}n Fuentes.
\newblock The extended evolutionary synthesis, ethnography, and the human
  niche: Toward an integrated anthropology.
\newblock {\em Current Anthropology}, 57(S13):S13--S26, 2016.

\bibitem{fuentes2017creative}
Agust{\'\i}n Fuentes.
\newblock {\em The creative spark: How imagination made humans exceptional}.
\newblock Penguin, 2017.

\bibitem{brown1991univ}
Donald~E. Brown.
\newblock {\em Human Universals}.
\newblock McGraw-Hill, 1991.

\bibitem{grinnell1917niche}
Joseph Grinnell.
\newblock The niche-relationships of the california thrasher.
\newblock {\em The Auk}, 34(4):427--433, 1917.

\bibitem{odum1971fundamentals}
Eugene~T. Odum.
\newblock {\em Fundamentals of Ecology}.
\newblock W.B. Saunders Company, 3rd edition, 1971.

\bibitem{peterson2003predicting}
A~Townsend Peterson.
\newblock Predicting the geography of species’ invasions via ecological niche
  modeling.
\newblock {\em The quarterly review of biology}, 78(4):419--433, 2003.

\bibitem{banks2006eco}
William~E Banks, FrancEsco d’Errico, Harold~L Dibble, Leonard Krishtalka,
  Dixie West, Deborah~I Olszewski, A~Townsend Peterson, David~G Anderson,
  JC~Gillam, Anta Montet-White, et~al.
\newblock Eco-cultural niche modeling: new tools for reconstructing the
  geography and ecology of past human populations.
\newblock {\em PaleoAnthropology}, 4(6):68--83, 2006.

\bibitem{elton2001animal}
Charles~S Elton.
\newblock {\em Animal Ecology}.
\newblock University of Chicago Press, 1927.

\bibitem{colwell2009hutchinson}
Robert~K Colwell and Thiago~F Rangel.
\newblock Hutchinson's duality: the once and future niche.
\newblock {\em Proceedings of the National Academy of Sciences},
  106(supplement\_2):19651--19658, 2009.

\bibitem{hutchinson1957multivariate}
G~Evelyn Hutchinson.
\newblock Concluding remarks.
\newblock In {\em Cold Spring Harbor Symposia on Quantitative Biology},
  volume~22, pages 415--421, 1957.

\bibitem{blonder2018new}
Benjamin Blonder, Cecina~Babich Morrow, Brian Maitner, David~J Harris,
  Christine Lamanna, Cyrille Violle, Brian~J Enquist, and Andrew~J Kerkhoff.
\newblock New approaches for delineating n-dimensional hypervolumes.
\newblock {\em Methods in Ecology and Evolution}, 9(2):305--319, 2018.

\bibitem{holt2009bringing}
Robert~D Holt.
\newblock Bringing the hutchinsonian niche into the 21st century: ecological
  and evolutionary perspectives.
\newblock {\em Proceedings of the National Academy of Sciences},
  106(supplement\_2):19659--19665, 2009.

\bibitem{hardesty1975niche}
Donald~L Hardesty.
\newblock The niche concept: suggestions for its use in human ecology.
\newblock {\em Human Ecology}, 3(2):71--85, 1975.

\bibitem{hardesty1977ecological}
Donald~L Hardesty.
\newblock {\em Ecological Anthropology}.
\newblock John Wiley adn Sons, 1977.

\bibitem{soberon2017fundamental}
Jorge Sober{\'o}n and B~Arroyo-Pe{\~n}a.
\newblock Are fundamental niches larger than the realized? testing a
  50-year-old prediction by hutchinson.
\newblock {\em Plos one}, 12(4):e0175138, 2017.

\bibitem{banks2008human}
William~E Banks, Francesco d'Errico, A~Townsend Peterson, Marian Vanhaeren,
  Masa Kageyama, Pierre Sepulchre, Gilles Ramstein, Anne Jost, and Daniel Lunt.
\newblock Human ecological niches and ranges during the lgm in europe derived
  from an application of eco-cultural niche modeling.
\newblock {\em Journal of Archaeological Science}, 35(2):481--491, 2008.

\bibitem{banks2013human}
William~E Banks, Francesco d'Errico, and Jo{\~a}o Zilh{\~a}o.
\newblock Human--climate interaction during the early upper paleolithic:
  testing the hypothesis of an adaptive shift between the proto-aurignacian and
  the early aurignacian.
\newblock {\em Journal of Human Evolution}, 64(1):39--55, 2013.

\bibitem{jones1994organisms}
Clive~G Jones, John~H Lawton, and Moshe Shachak.
\newblock Organisms as ecosystem engineers.
\newblock {\em Oikos}, 69(3):373--386, 1994.

\bibitem{jones1997positive}
Clive~G Jones, John~H Lawton, and Moshe Shachak.
\newblock Positive and negative effects of organisms as physical ecosystem
  engineers.
\newblock {\em Ecology}, 78(7):1946--1957, 1997.

\bibitem{smith2007ultimate}
Bruce~D Smith.
\newblock The ultimate ecosystem engineers.
\newblock {\em Science}, 315(5820):1797--1798, 2007.

\bibitem{faith2021rethinking}
J~Tyler Faith, Andrew Du, Anna~K Behrensmeyer, Benjamin Davies, David~B
  Patterson, John Rowan, and Bernard Wood.
\newblock Rethinking the ecological drivers of hominin evolution.
\newblock {\em Trends in Ecology \& Evolution}, 36(9):797--807, 2021.

\bibitem{braun2021ecosystem}
David~R Braun, John~Tyler Faith, Matthew~J Douglass, Benjamin Davies, Mitchel~J
  Power, Vera Aldeias, Nicholas~J Conard, Russell Cutts, Larisa~RG DeSantis,
  Lydie~M Dupont, et~al.
\newblock Ecosystem engineering in the quaternary of the west coast of south
  africa.
\newblock {\em Evolutionary Anthropology: Issues, News, and Reviews},
  30(1):50--62, 2021.

\bibitem{wrangham2009catching}
Richard Wrangham.
\newblock {\em Catching fire: how cooking made us human}.
\newblock Basic books, 2009.

\bibitem{odling2013niche}
F~John Odling-Smee, Kevin~N Laland, and Marcus~W Feldman.
\newblock {\em Niche Construction}.
\newblock Princeton University Press, 2003.

\bibitem{barker2014integrating}
Gillian Barker and John Odling-Smee.
\newblock Integrating ecology and evolution: Niche construction and ecological
  engineering.
\newblock In {\em Entangled life}, pages 187--211. Springer, 2014.

\bibitem{laland2001cultural}
Kevin~N Laland, John Odling-Smee, and Marcus~W Feldman.
\newblock Cultural niche construction and human evolution.
\newblock {\em Journal of evolutionary biology}, 14(1):22--33, 2001.

\bibitem{laland2010niche}
Kevin~N Laland and Michael~J O’Brien.
\newblock Niche construction theory and archaeology.
\newblock {\em Journal of Archaeological Method and Theory}, 17(4):303--322,
  2010.

\bibitem{boivin2016ecological}
Nicole~L Boivin, Melinda~A Zeder, Dorian~Q Fuller, Alison Crowther, Greger
  Larson, Jon~M Erlandson, Tim Denham, and Michael~D Petraglia.
\newblock Ecological consequences of human niche construction: Examining
  long-term anthropogenic shaping of global species distributions.
\newblock {\em Proceedings of the National Academy of Sciences},
  113(23):6388--6396, 2016.

\bibitem{gerbault2011evolution}
Pascale Gerbault, Anke Liebert, Yuval Itan, Adam Powell, Mathias Currat,
  Joachim Burger, Dallas~M Swallow, and Mark~G Thomas.
\newblock Evolution of lactase persistence: an example of human niche
  construction.
\newblock {\em Philosophical Transactions of the Royal Society B: Biological
  Sciences}, 366(1566):863--877, 2011.

\bibitem{obrien2012genes}
Michael~J O’Brien and Kevin~N Laland.
\newblock Genes, culture, and agriculture: An example of human niche
  construction.
\newblock {\em Current Anthropology}, 53(4):434--470, 2012.

\bibitem{bird2013niche}
Rebecca~Bliege Bird, Nyalangka Tayor, Brian~F Codding, and Douglas~W Bird.
\newblock Niche construction and dreaming logic: aboriginal patch mosaic
  burning and varanid lizards (varanus gouldii) in australia.
\newblock {\em Proceedings of the Royal Society B: Biological Sciences},
  280(1772):20132297, 2013.

\bibitem{obrien2021genes}
Michael~J O'Brien and R~Alexander Bentley.
\newblock Genes, culture, and the human niche: An overview.
\newblock {\em Evolutionary Anthropology: Issues, News, and Reviews},
  30(1):40--49, 2021.

\bibitem{brown2004toward}
James~H Brown, James~F Gillooly, Andrew~P Allen, Van~M Savage, and Geoffrey~B
  West.
\newblock Toward a metabolic theory of ecology.
\newblock {\em Ecology}, 85(7):1771--1789, 2004.

\bibitem{haskell2002fractal}
John~P Haskell, Mark~E Ritchie, and Han Olff.
\newblock Fractal geometry predicts varying body size scaling relationships for
  mammal and bird home ranges.
\newblock {\em Nature}, 418(6897):527--530, 2002.

\bibitem{ritchie2009scale}
Mark~E Ritchie.
\newblock {\em Scale, Heterogeneity, and the Structure and Diversity of
  Ecological Communities}.
\newblock Princeton University Press, 2009.

\bibitem{wilson1975adequacy}
David~Sloan Wilson.
\newblock The adequacy of body size as a niche difference.
\newblock {\em The American Naturalist}, 109(970):769--784, 1975.

\bibitem{woodward2005body}
Guy Woodward, Bo~Ebenman, Mark Emmerson, Jose~M Montoya, Jens~M Olesen, Alfredo
  Valido, and Philip~H Warren.
\newblock Body size in ecological networks.
\newblock {\em Trends in ecology \& evolution}, 20(7):402--409, 2005.

\bibitem{pontzer2015energy}
Herman Pontzer, David~A Raichlen, Brian~M Wood, Melissa Emery~Thompson, Susan~B
  Racette, Audax~ZP Mabulla, and Frank~W Marlowe.
\newblock Energy expenditure and activity among hadza hunter-gatherers.
\newblock {\em American Journal of Human Biology}, 27(5):628--637, 2015.

\bibitem{peterson1990sustained}
Charles~C Peterson, Kenneth~A Nagy, and Jared Diamond.
\newblock Sustained metabolic scope.
\newblock {\em Proceedings of the National Academy of Sciences},
  87(6):2324--2328, 1990.

\bibitem{brown2011energetic}
James~H Brown, William~R Burnside, Ana~D Davidson, John~P DeLong, William~C
  Dunn, Marcus~J Hamilton, Norman Mercado-Silva, Jeffrey~C Nekola, Jordan~G
  Okie, William~H Woodruff, et~al.
\newblock Energetic limits to economic growth.
\newblock {\em BioScience}, 61(1):19--26, 2011.

\bibitem{burger2011industrial}
Oskar Burger, John~P DeLong, and Marcus~J Hamilton.
\newblock Industrial energy use and the human life history.
\newblock {\em Scientific reports}, 1(1):1--7, 2011.

\bibitem{van1973new}
Leigh Van~Valen.
\newblock A new evolutionary law.
\newblock {\em Evol theory}, 1:1--30, 1973.

\bibitem{tomasello1993cultural}
Michael Tomasello, Ann~Cale Kruger, and Hilary~Horn Ratner.
\newblock Cultural learning.
\newblock {\em Behavioral and brain sciences}, 16(3):495--511, 1993.

\bibitem{mesoudi2018cumulative}
Alex Mesoudi and Alex Thornton.
\newblock What is cumulative cultural evolution?
\newblock {\em Proceedings of the Royal Society B}, 285(1880):20180712, 2018.

\bibitem{pinker2010cognitive}
Steven Pinker.
\newblock The cognitive niche: Coevolution of intelligence, sociality, and
  language.
\newblock {\em Proceedings of the National Academy of Sciences},
  107(supplement\_2):8993--8999, 2010.

\bibitem{whiten2012human}
Andrew Whiten and David Erdal.
\newblock The human socio-cognitive niche and its evolutionary origins.
\newblock {\em Philosophical Transactions of the Royal Society B: Biological
  Sciences}, 367(1599):2119--2129, 2012.

\bibitem{tomasello2005understanding}
Michael Tomasello, Malinda Carpenter, Josep Call, Tanya Behne, and Henrike
  Moll.
\newblock Understanding and sharing intentions: The origins of cultural
  cognition.
\newblock {\em Behavioral and brain sciences}, 28(5):675--691, 2005.

\bibitem{herculano2015mammalian}
Suzana Herculano-Houzel, Kenneth Catania, Paul~R Manger, and Jon~H Kaas.
\newblock Mammalian brains are made of these: a dataset of the numbers and
  densities of neuronal and nonneuronal cells in the brain of glires, primates,
  scandentia, eulipotyphlans, afrotherians and artiodactyls, and their
  relationship with body mass.
\newblock {\em Brain, Behavior and Evolution}, 86(3-4):145--163, 2015.

\bibitem{herculano2017numbers}
Suzana Herculano-Houzel.
\newblock Numbers of neurons as biological correlates of cognitive capability.
\newblock {\em Current Opinion in Behavioral Sciences}, 16:1--7, 2017.

\bibitem{el2011network}
Abbas El~Gamal and Young-Han Kim.
\newblock {\em Network information theory}.
\newblock Cambridge university press, 2011.

\bibitem{stone2018principles}
James~V Stone.
\newblock {\em Principles of neural information theory}.
\newblock Sebtel Press, 2018.

\bibitem{martin1981relative}
Robert~D Martin.
\newblock Relative brain size and basal metabolic rate in terrestrial
  vertebrates.
\newblock {\em Nature}, 293(5827):57--60, 1981.

\bibitem{burger2019allometry}
Joseph~Robert Burger, Menshian~Ashaki George~Jr, Claire Leadbetter, and Farhin
  Shaikh.
\newblock The allometry of brain size in mammals.
\newblock {\em Journal of Mammalogy}, 100(2):276--283, 2019.

\bibitem{jerison2012evolution}
Harry Jerison.
\newblock {\em Evolution of the brain and intelligence}.
\newblock Elsevier, 1973.

\bibitem{harvey1991comparative}
Paul~H Harvey, Mark~D Pagel, et~al.
\newblock {\em The comparative method in evolutionary biology}, volume 239.
\newblock Oxford university press Oxford, 1991.

\bibitem{deaner2007overall}
Robert~O Deaner, Karin Isler, Judith Burkart, and Carel Van~Schaik.
\newblock Overall brain size, and not encephalization quotient, best predicts
  cognitive ability across non-human primates.
\newblock {\em Brain, behavior and evolution}, 70(2):115--124, 2007.

\bibitem{azevedo2009equal}
Frederico~AC Azevedo, Ludmila~RB Carvalho, Lea~T Grinberg, Jos{\'e}~Marcelo
  Farfel, Renata~EL Ferretti, Renata~EP Leite, Wilson~Jacob Filho, Roberto
  Lent, and Suzana Herculano-Houzel.
\newblock Equal numbers of neuronal and nonneuronal cells make the human brain
  an isometrically scaled-up primate brain.
\newblock {\em Journal of Comparative Neurology}, 513(5):532--541, 2009.

\bibitem{herculano2012remarkable}
Suzana Herculano-Houzel.
\newblock The remarkable, yet not extraordinary, human brain as a scaled-up
  primate brain and its associated cost.
\newblock {\em Proceedings of the National Academy of Sciences},
  109(supplement\_1):10661--10668, 2012.

\bibitem{passingham2021understanding}
Richard Passingham.
\newblock {\em Understanding the prefrontal cortex: selective advantage,
  connectivity, and neural operations}.
\newblock Oxford University Press, 2021.

\bibitem{hamilton2022collective}
Marcus~J Hamilton.
\newblock Collective computation, information flow, and the emergence of
  hunter-gatherer small-worlds.
\newblock {\em Journal of Social Computing}, 3(1):18--37, 2022.

\bibitem{passingham2008special}
Richard Passingham.
\newblock {\em What is special about the human brain?}
\newblock Oxford University Press, 2008.

\bibitem{dunbar1992neocortex}
Robin~IM Dunbar.
\newblock Neocortex size as a constraint on group size in primates.
\newblock {\em Journal of human evolution}, 22(6):469--493, 1992.

\bibitem{dunbar1998social}
Robin~IM Dunbar.
\newblock The social brain hypothesis.
\newblock {\em Evolutionary Anthropology: Issues, News, and Reviews: Issues,
  News, and Reviews}, 6(5):178--190, 1998.

\bibitem{trivers1971evolution}
Robert~L Trivers.
\newblock The evolution of reciprocal altruism.
\newblock {\em The Quarterly Review of Biology}, 46(1):35--57, 1971.

\bibitem{lancaster2017watershed}
Jane~B Lancaster and Chet~S Lancaster.
\newblock The watershed: Change in parental-investment and family-formation
  strategies in the course of human evolution.
\newblock In {\em Parenting across the life span}, pages 187--206. Routledge,
  1987.

\bibitem{hamilton2007nonlinear}
Marcus~J Hamilton, Bruce~T Milne, Robert~S Walker, and James~H Brown.
\newblock Nonlinear scaling of space use in human hunter--gatherers.
\newblock {\em Proceedings of the National Academy of Sciences},
  104(11):4765--4769, 2007.

\bibitem{shipton2015before}
Ceri Shipton and Mark Nielsen.
\newblock Before cumulative culture.
\newblock {\em Human Nature}, 26(3):331--345, 2015.

\bibitem{tennie2009ratcheting}
Claudio Tennie, Josep Call, and Michael Tomasello.
\newblock Ratcheting up the ratchet: on the evolution of cumulative culture.
\newblock {\em Philosophical Transactions of the Royal Society B: Biological
  Sciences}, 364(1528):2405--2415, 2009.

\bibitem{hamilton2007complex}
Marcus~J Hamilton, Bruce~T Milne, Robert~S Walker, Oskar Burger, and James~H
  Brown.
\newblock The complex structure of hunter--gatherer social networks.
\newblock {\em Proceedings of the Royal Society B: Biological Sciences},
  274(1622):2195--2203, 2007.

\bibitem{granovetter1973strength}
Mark~S Granovetter.
\newblock The strength of weak ties.
\newblock {\em American journal of sociology}, 78(6):1360--1380, 1973.

\bibitem{zhou2005discrete}
W-X Zhou, Didier Sornette, Russell~A Hill, and Robin~IM Dunbar.
\newblock Discrete hierarchical organization of social group sizes.
\newblock {\em Proceedings of the Royal Society B: Biological Sciences},
  272(1561):439--444, 2005.

\bibitem{fuchs2014fractal}
Benedikt Fuchs, Didier Sornette, and Stefan Thurner.
\newblock Fractal multi-level organisation of human groups in a virtual world.
\newblock {\em Scientific reports}, 4(1):1--6, 2014.

\bibitem{hamilton2020scaling}
Marcus~J Hamilton, Robert~S Walker, Briggs Buchanan, and David~S Sandeford.
\newblock Scaling human sociopolitical complexity.
\newblock {\em PloS one}, 15(7):e0234615, 2020.

\bibitem{brush2018conflicts}
Eleanor~R Brush, David~C Krakauer, and Jessica~C Flack.
\newblock Conflicts of interest improve collective computation of adaptive
  social structures.
\newblock {\em Science advances}, 4(1):e1603311, 2018.

\bibitem{anderson1972more}
Philip~W Anderson.
\newblock More is different: broken symmetry and the nature of the hierarchical
  structure of science.
\newblock {\em Science}, 177(4047):393--396, 1972.

\bibitem{west2018scale}
Geoffrey West.
\newblock {\em Scale: The universal laws of life, growth, and death in
  organisms, cities, and companies}.
\newblock Penguin, 2018.

\bibitem{brown2014there}
James~H Brown.
\newblock Why are there so many species in the tropics?
\newblock {\em Journal of biogeography}, 41(1):8--22, 2014.

\bibitem{hamilton2020diversity}
Marcus~J Hamilton, Robert~S Walker, and Christopher~P Kempes.
\newblock Diversity begets diversity in mammal species and human cultures.
\newblock {\em Scientific reports}, 10(1):19654, 2020.

\bibitem{freeman2016socioecology}
Jacob Freeman.
\newblock The socioecology of territory size and a" work-around" hypothesis for
  the adoption of farming.
\newblock {\em PloS one}, 11(7):e0158743, 2016.

\bibitem{walker2014remote}
Robert~S Walker, Marcus~J Hamilton, and Aaron~A Groth.
\newblock Remote sensing and conservation of isolated indigenous villages in
  amazonia.
\newblock {\em Royal Society open science}, 1(3):140246, 2014.

\bibitem{hamilton2012human}
Marcus~J Hamilton, Oskar Burger, and Robert~S Walker.
\newblock Human ecology.
\newblock {\em Metabolic ecology: a scaling approach}, pages 248--257, 2012.

\bibitem{bettencourt2007growth}
Lu{\'\i}s~MA Bettencourt, Jos{\'e} Lobo, Dirk Helbing, Christian K{\"u}hnert,
  and Geoffrey~B West.
\newblock Growth, innovation, scaling, and the pace of life in cities.
\newblock {\em Proceedings of the Ntional Academy of Sciences},
  104(17):7301--7306, 2007.

\bibitem{burger2017extra}
Joseph~R Burger, Vanessa~P Weinberger, and Pablo~A Marquet.
\newblock Extra-metabolic energy use and the rise in human hyper-density.
\newblock {\em Scientific reports}, 7(1):43869, 2017.

\bibitem{morris2013measure}
Ian Morris.
\newblock {\em The measure of civilization}.
\newblock Princeton University Press, 2013.

\bibitem{bettencourt2010unified}
Luis Bettencourt and Geoffrey West.
\newblock A unified theory of urban living.
\newblock {\em Nature}, 467(7318):912--913, 2010.

\bibitem{crutzen2006anthropocene}
Paul~J Crutzen.
\newblock The “anthropocene”.
\newblock In {\em Earth System Science in the Anthropocene}, pages 13--18.
  Springer, 2006.

\bibitem{smil2018energy}
Vaclav Smil.
\newblock {\em Energy and civilization: a history}.
\newblock MIT Press, 2018.

\bibitem{brown2014macroecology}
James~H Brown, Joseph~R Burger, William~R Burnside, Michael Chang, Ana~D
  Davidson, Trevor~S Fristoe, Marcus~J Hamilton, Sean~T Hammond, Astrid
  Kodric-Brown, Norman Mercado-Silva, et~al.
\newblock Macroecology meets macroeconomics: Resource scarcity and global
  sustainability.
\newblock {\em Ecological engineering}, 65:24--32, 2014.

\bibitem{bettencourt2013origins}
Lu{\'\i}s~MA Bettencourt.
\newblock The origins of scaling in cities.
\newblock {\em Science}, 340(6139):1438--1441, 2013.

\bibitem{ortman2014pre}
Scott~G Ortman, Andrew~HF Cabaniss, Jennie~O Sturm, and Lu{\'\i}s~MA
  Bettencourt.
\newblock The pre-history of urban scaling.
\newblock {\em PloS one}, 9(2):e87902, 2014.

\bibitem{ortman2015settlement}
Scott~G Ortman, Andrew~HF Cabaniss, Jennie~O Sturm, and Lu{\'\i}s~MA
  Bettencourt.
\newblock Settlement scaling and increasing returns in an ancient society.
\newblock {\em Science Advances}, 1(1):e1400066, 2015.

\bibitem{lobo2020urban}
Jos{\'e} Lobo, Marina Alberti, Melissa Allen-Dumas, Elsa Arcaute, Marc
  Barthelemy, Luis~A Bojorquez~Tapia, Shauna Brail, Luis Bettencourt, Anni
  Beukes, Wei-Qiang Chen, et~al.
\newblock Urban science: Integrated theory from the first cities to sustainable
  metropolises.
\newblock {\em arxiv}, 2020.

\bibitem{hamilton2018scaling}
Marcus~J Hamilton, Briggs Buchanan, and Robert~S Walker.
\newblock Scaling the size, structure, and dynamics of residentially mobile
  hunter-gatherer camps.
\newblock {\em American Antiquity}, 83(4):701--720, 2018.

\bibitem{lobo2022scaling}
Jos{\'e} Lobo, Todd Whitelaw, Lu{\'\i}s~MA Bettencourt, Polly Wiessner,
  Michael~E Smith, and Scott Ortman.
\newblock Scaling of hunter-gatherer camp size and human sociality.
\newblock {\em Current Anthropology}, 63(1):68--94, 2022.

\bibitem{haas2015settlement}
W~Randall Haas~Jr, Cynthia~J Klink, Greg~J Maggard, and Mark~S Aldenderfer.
\newblock Settlement-size scaling among prehistoric hunter-gatherer settlement
  systems in the new world.
\newblock {\em PloS one}, 10(11):e0140127, 2015.

\bibitem{derex2019causal}
Maxime Derex, Jean-Fran{\c{c}}ois Bonnefon, Robert Boyd, and Alex Mesoudi.
\newblock Causal understanding is not necessary for the improvement of
  culturally evolving technology.
\newblock {\em Nature human behaviour}, 3(5):446--452, 2019.

\bibitem{wolpertArxiv}
David~H. Wolpert.
\newblock What can we know about that which we cannot even imagine?, 2022.

\bibitem{bostrom2017superintelligence}
Nick Bostrom.
\newblock {\em Superintelligence}.
\newblock Dunod, 2017.

\bibitem{mitchell2019artificial}
Melanie Mitchell.
\newblock {\em Artificial intelligence: A guide for thinking humans}.
\newblock Penguin UK, 2019.

\bibitem{tegmark2018life}
Max Tegmark.
\newblock {\em Life 3.0: Being human in the age of artificial intelligence}.
\newblock Vintage, 2018.

\bibitem{preskill2018quantum}
John Preskill.
\newblock Quantum computing in the nisq era and beyond.
\newblock {\em Quantum}, 2:79, 2018.

\bibitem{doe2023}
Department of~Energy.
\newblock Doe national laboratory makes history by achieving fusion ignition.
\newblock {\em DOE National Laboratory Makes History by Achieving Fusion
  Ignition}, Dec 2023.

\bibitem{kurzweil2014singularity}
Ray Kurzweil.
\newblock {\em The singularity is near}.
\newblock Springer, 2014.

\bibitem{siddiqi2018beyond}
Asif~A Siddiqi.
\newblock {\em Beyond Earth: A chronicle of deep space exploration, 1958-2016},
  volume 4041.
\newblock National Aeronautis \& Space Administration, 2018.

\bibitem{white2001humans}
Ronald~J White and Maurice Averner.
\newblock Humans in space.
\newblock {\em Nature}, 409(6823):1115--1118, 2001.

\bibitem{levchenko2021mars}
Igor Levchenko, Shuyan Xu, St{\'e}phane Mazouffre, Michael Keidar, and Kateryna
  Bazaka.
\newblock Mars colonization: beyond getting there.
\newblock {\em Terraforming Mars}, pages 73--98, 2021.

\bibitem{goldsmith2022end}
Donald Goldsmith and Martin Rees.
\newblock {\em The End of Astronauts}.
\newblock Harvard University Press, 2022.

\bibitem{afshinnekoo2020fundamental}
Ebrahim Afshinnekoo, Ryan~T Scott, Matthew~J MacKay, Eloise Pariset, Egle
  Cekanaviciute, Richard Barker, Simon Gilroy, Duane Hassane, Scott~M Smith,
  Sara~R Zwart, et~al.
\newblock Fundamental biological features of spaceflight: advancing the field
  to enable deep-space exploration.
\newblock {\em Cell}, 183(5):1162--1184, 2020.

\bibitem{gunga2020human}
Hanns-Christian Gunga.
\newblock Space.
\newblock In {\em Human Physiology in Extreme Environments}, pages 285--335.
  Academic Press, 2020.

\bibitem{cahill2020space}
T.~Cahill and G.~Hardiman.
\newblock Nutritional challenges and countermeasures for space travel.
\newblock {\em Nutrition Bulletin}, 45(1):98--105, 2020.

\bibitem{wheeler2001plant}
Raymond~M Wheeler, Gary~W Stutte, GV~Subbarao, and Neil~C Yorio.
\newblock Plant growth and human life support for space travel.
\newblock In {\em Handbook of plant and crop physiology}, pages 925--941. CRC
  Press, 2001.

\bibitem{kanas2008space}
Nick Kanas and Dietrich Manzey.
\newblock {\em Space psychology and psychiatry}, volume~16.
\newblock Springer, 2008.

\bibitem{durante2011physical}
Marco Durante and Francis~A Cucinotta.
\newblock Physical basis of radiation protection in space travel.
\newblock {\em Reviews of modern physics}, 83(4):1245, 2011.

\end{thebibliography}

\end{document}